\documentclass[12pt]{article}
  \usepackage{amsfonts}
  \usepackage{amsmath}
\usepackage{amssymb}
\usepackage{amscd}
\usepackage{graphicx}
\begin{document}

~~
\bigskip
\bigskip
\begin{center}
{\Large {\bf{{{Chaos synchronization of identical Sprott systems by
active control}}}}}
\end{center}
\bigskip
\bigskip
\bigskip
\begin{center}
{{\large ${\rm {Marcin\;Daszkiewicz}}$}}
\end{center}
\bigskip
\begin{center}
\bigskip

{ ${\rm{Institute\; of\; Theoretical\; Physics}}$}

{ ${\rm{ University\; of\; Wroclaw\; pl.\; Maxa\; Borna\; 9,\;
50-206\; Wroclaw,\; Poland}}$}

{ ${\rm{ e-mail:\; marcin@ift.uni.wroc.pl}}$}

\end{center}
\bigskip
\bigskip
\bigskip
\bigskip
\bigskip
\bigskip
\bigskip
\bigskip
\bigskip
\begin{abstract}
In this article we synchronize by active control method all 19 identical Sprott systems
provided in paper \cite{[10]}. Particularly, we find the corresponding active
controllers as well as we perform (as an example) the numerical synchronization
of two Sprott-A models.
\end{abstract}
\bigskip
\bigskip
\bigskip
\bigskip
\eject

\section{{{Introduction}}}

In the last four decades there appeared a lot of papers dealing with so-called chaotic
models, i.e., with such models whose dynamics is described by strongly sensitive with
respect initial conditions, nonlinear differential equations. The most popular of them
are: Lorenz system \cite{[1]}, Roessler system \cite{[2]}, Rayleigh-Benard system \cite{[3]}, Henon-Heiles
system \cite{[4]}, jerk equation \cite{[5]}, Duffing equation \cite{[6]}, Lotka-Volter system \cite{[7]}, Liu system
\cite{[8]}, Chen system \cite{[9]} and Sprott system \cite{[10]}. A lot of them have been applied in various
fields of industrial and scientific divisions, such as, for example: Physics, Chemistry,
Biology, Microbiology, Economics, Electronics, Engineering, Computer Science, Secure
Communications, Image Processing and Robotics.

One of the most important problem of the chaos theory concerns so-called chaos synchronization
phenomena. Since Pecora and Caroll \cite{[11]} introduced a method to synchronize
two identical chaotic systems, the chaos synchronization has received increasing attention
due to great potential applications in many scientific discipline. Generally, there are
known several methods of chaos synchronization, such as: OGY method \cite{[12]}, active con-
trol method \cite{[13]}-\cite{[16]}, adaptive control method \cite{[17]}-\cite{[21]}, backstepping method \cite{[22]}, \cite{[23]},
sampled-data feedback synchronization method \cite{[24]}, time-delay feedback method \cite{[25]} and
sliding mode control method \cite{[26]}-\cite{[29]}.

In this article we synchronize by active control scheme all 19 identical Sprott systems
provided in publication \cite{[10]}. Particularly, we establish the proper so-called active con-
trollers with use of the Lyapunov stabilization theory \cite{[30]}. It should be noted, however,
that some types of Sprott model have been already synchronized with use of the active
control method in papers \cite{[31]} and \cite{[32]}. Apart of that there has been synchronized the
Sprott systems in the framework of adaptive control scheme in articles \cite{[33]} and \cite{[34]}.

The paper is organized as follows. In second Section we recall the main result of Sprott
article \cite{[10]}, i.e., we provide Table 1 including dynamics of all 19 Sprott chaotic systems.
In Section 3 we remaind the basic concepts of active synchronization method, while in
Section 4 we consider as an example the synchronization of two identical Sprott-A models.
The fifth Section is devoted to the main result of the paper, and it provides in Table 2
the active controllers which synchronize all identical Sprott systems. The conclusions and
final remarks are discussed in the last Section.

\section{Sprott systems}

In this Section we recall the main result of paper \cite{[10]}, in which there has been performed a
systematic examination of general three-dimensional ordinary differential equations with
quadratic nonlinearities. Particularly, it has been uncovered 19 distinct simple examples
of chaotic flows (so-called Sprott systems) listed in {\bf Table 1}.\\
\\
\\
\begin{tabular}{ccccc}
\hline $\;\;\;\;{\rm type}\;\;\;\;$ & $\;\;\;\;\;\;\;\;{\rm 1st\;equation}\;\;\;\;\;\;\;$ & $\;\;\;\;\;\;\;\;{\rm 2nd\;equation}\;\;\;\;\;\;\;$ & $\;\;\;\;\;\;\;\;{\rm 3rd\;equation}\;\;\;\;\;\;\;$\\
\hline ${\rm A}$ & $\dot{x}_1=x_2$ & $\dot{x}_2=-x_1+x_2x_3$ & $\dot{x}_3=1-x_2^2$\\
       ${\rm B}$ & $\dot{x}_1=x_2x_3$ & $\dot{x}_2=x_1-x_2$ & $\dot{x}_3=1-x_1x_2$\\
       ${\rm C}$ & $\dot{x}_1=x_2x_3$ & $\dot{x}_2=x_1-x_2$ & $\dot{x}_3=1-x_1^2$\\
       ${\rm D}$ & $\dot{x}_1=-x_2$ & $\dot{x}_2=x_1+x_3$ & $\dot{x}_3=x_1x_3+ 3x_2^2$\\
       ${\rm E}$ & $\dot{x}_1=x_2x_3$ & $\dot{x}_2=x_1^2-x_2$ & $\dot{x}_3=1-4x_1$\\
       ${\rm F}$ & $\dot{x}_1=x_2+x_3$ & $\dot{x}_2=-x_1+0.5x_2$ & $\dot{x}_3=x_1^2-x_3$\\
       ${\rm G}$ & $\dot{x}_1=0.4x_1+x_3$ & $\dot{x}_2=x_1x_3-x_2$ & $\dot{x}_3=-x_1+x_2$\\
       ${\rm H}$ & $\dot{x}_1=-x_2+x_3^2$ & $\dot{x}_2=x_1+0.5x_2$ & $\dot{x}_3=x_1-x_3$\\
       ${\rm I}$ & $\dot{x}_1=-0.2x_2$ & $\dot{x}_2=x_1+x_3$ & $\dot{x}_3=x_1+ x_2^2-x_3$\\
       ${\rm J}$ & $\dot{x}_1=2x_3$ & $\dot{x}_2=-2x_2+x_3$ & $\dot{x}_3=-x_1+x_2+x_2^2$\\
       ${\rm K}$ & $\dot{x}_1=x_1x_2 -x_3$ & $\dot{x}_2=x_1-x_2$ & $\dot{x}_3=x_1+0.3x_3$\\
       ${\rm L}$ & $\dot{x}_1=x_2+3.9x_3$ & $\dot{x}_2=0.9x_1^2-x_2$ & $\dot{x}_3=1-x_1$\\
       ${\rm M}$ & $\dot{x}_1=-x_3$ & $\dot{x}_2=-x_1^2-x_2$ & $\dot{x}_3=1.7+1.7x_1+x_2$\\
       ${\rm N}$ & $\dot{x}_1=-2x_2$ & $\dot{x}_2=x_1+x_3^2$ & $\dot{x}_3=1+x_2-2x_1$\\
       ${\rm O}$ & $\dot{x}_1=x_2$ & $\dot{x}_2=x_1-x_3$ & $\dot{x}_3=x_1+x_1x_3+2.7x_2$\\
       ${\rm P}$ & $\dot{x}_1=2.7x_2+x_3$ & $\dot{x}_2=-x_1+x_2^2$ & $\dot{x}_3=x_1+x_2$\\
       ${\rm Q}$ & $\dot{x}_1=-x_3$ & $\dot{x}_2=x_1-x_2$ & $\dot{x}_3=3.1x_1+x_2^2+0.5x_3$\\
       ${\rm R}$ & $\dot{x}_1=0.9-x_2$ & $\dot{x}_2=0.4+x_3$ & $\dot{x}_3=x_1x_2-x_3$\\
       ${\rm S}$ & $\dot{x}_1=-x_1-4x_2$ & $\dot{x}_2=x_1+x_3^2$ & $\dot{x}_3=1+x_1$\\
\hline
\end{tabular}

\begin{center}
{\bf Table 1.} The Sprott systems.
\end{center}

\section{Chaos synchronization by active control - general prescription}

In this Section we remaind the general scheme of chaos synchronization of two systems
by so-called active control procedure \cite{[13]}-\cite{[16]}.
Let us start with the following master model\footnote{$\frac{do}{dt} = \dot{o}$.}
\begin{eqnarray}
\dot{x} = Ax + F(x) \;, \label{eq1}
\end{eqnarray}
where $x = [\;x_1, x_2,\dots ,x_n\;]$ is the state of the system, A denotes the $n \times n$ matrix of
the system parameters and $F(x)$ plays the role of the nonlinear part of the differential
equation (\ref{eq1}). The slave model dynamics is described by
\begin{eqnarray}
\dot{y} = By + G(y) + u \;,\label{eq2}
\end{eqnarray}
with $y = [\;y_1, y_2,\dots ,y_n\;]$ being the state of the system, $B$ denoting the $n$-dimensional
quadratic matrix of the system, $G(y)$ playing the role of nonlinearity of the equation (\ref{eq2})
and $u = [\;u_1, u_2,\dots ,u_n\;]$ being the active controller of the slave model. Besides, it should
be mentioned that for matrices $A = B$ and functions $F = G$ the states $x$ and $y$ describe
two identical chaotic systems. In the case $A \neq B$ or $F \neq G$ they correspond to the two
different chaotic models.

Let us now provide the following synchronization error vector
\begin{eqnarray}
e = y - x \;, \label{eq3}
\end{eqnarray}
which in accordance with  (\ref{eq1}) and  (\ref{eq2}) obeys
\begin{eqnarray}
\dot{e} = By - Ax + G(y) - F(x) + u\;. \label{eq4}
\end{eqnarray}

In active control method we try to find such a controller $u$, which synchronizes the state
of the master system (\ref{eq1}) with the state of the slave system (\ref{eq2}) for any initial condition
$x_0 = x(0)$ and $y_0 = y(0)$. In other words we design a controller $u$ in such a way that for
system (\ref{eq4}) we have
\begin{eqnarray}
\lim_{t \to \infty}||{e}(t)|| =0\;, \label{eq5}
\end{eqnarray}
for all initial conditions $e_0 = e(0)$. In order to establish the synchronization (\ref{eq4}) we use
the Lyapunov stabilization theory \cite{[30]}. It means, that if we take as a candidate Lyapunov
function of the form
\begin{eqnarray}
V(e) = e^{T}PV(e)e \;, \label{eq6}
\end{eqnarray}
with $P$ being a positive $n \times n$ matrix, then we wish to find the active controller $u$ so that
\begin{eqnarray}
\dot{V}(e) = -e^{T}QV(e)e \;, \label{eq7}
\end{eqnarray}
where Q is a positive definite $n \times n$ matrix as well. Then the systems (\ref{eq1}) and (\ref{eq2}) remain
synchronized.

\section{The example: chaos synchronization of (identical) Sprott-A systems}

In accordance with two pervious Sections the master Sprott-A system is described by the
following dynamics (see {\bf Table 1})
\begin{equation}
\left\{\begin{array}{rcl}
\dot{x}_1 &=& x_2\;\\[5pt]
&~~&~\cr
\dot{x}_2 &=& -x_1 + x_2x_3\;\\[5pt]
&~~&~\cr
\dot{x}_3 &=& 1-x_2^2\;,
\label{eq8}\end{array}\right.
\end{equation}
where functions $x_1$, $x_2$ and $x_3$ denote the states of the system; its slave Sprott-A partner
is given by
\begin{equation}
\left\{\begin{array}{rcl}
\dot{y}_1 &=& y_2 + u_1\;\\[5pt]
&~~&~\cr
\dot{y}_2 &=& -y_1 + y_2y_3 + u_2\;\\[5pt]
&~~&~\cr
\dot{y}_3 &=& 1-y_2^2 + u_3\;,
\label{eq9}\end{array}\right.
\end{equation}
with active controllers $u_1$, $u_2$ and $u_3$ respectively. Using (\ref{eq8}) and (\ref{eq9}) one can check that
the dynamics of synchronization errors $e_i = y_i - x_i$ is obtained as\footnote{See also formula (\ref{eq4}).}
\begin{equation}
\left\{\begin{array}{rcl}
\dot{e}_1 &=& e_2 + u_1\;\\[5pt]
&~~&~\cr
\dot{e}_2 &=& -e_1 + y_2y_3 - x_2x_3 + u_2\;\\[5pt]
&~~&~\cr
\dot{e}_3 &=& -e_2(y_2 + x_2) + u_3\;.
\label{eq10}\end{array}\right.
\end{equation}
Besides, if we define the positive Lyapunov function by\footnote{The matrix $P = 1$ in the formula (\ref{eq6}).}
\begin{eqnarray}
{V}(e) = \frac{1}{2}\left(e_1^2+e_2^2+e_3^2\right) \;, \label{eq11}
\end{eqnarray}
then for the following choice of control functions
\begin{equation}
\left\{\begin{array}{rcl}
{u}_1 &=& -(e_1+e_2) \;\\[5pt]
&~~&~\cr
{u}_2 &=& e_1 -e_2 - y_2y_3 + x_2x_3\;\\[5pt]
&~~&~\cr
{u}_3 &=& e_2(y_2 + x_2) - e_3\;,
\label{eq12}\end{array}\right.
\end{equation}
we have\footnote{The matrix $Q = 1$ in the formula (\ref{eq7}).}
\begin{eqnarray}
\dot{V}(e) = -\left(e_1^2+e_2^2+e_3^2\right) \;. \label{eq13}
\end{eqnarray}
Such a result means (see general prescription) that the identical Sprott-A systems (\ref{eq8}) and
(\ref{eq9}) are synchronized for all initial conditions with active controllers (\ref{eq12}).

Let us now illustrate the above considerations by the proper numerical calculations.

First of all, we solve the Sprott-A system with two different sets of initial conditions
\begin{eqnarray}
(x_{01},x_{02},x_{03}) = (1,0.05,0.05)\;, \label{eq14}
\end{eqnarray}
and
\begin{eqnarray}
(y_{01},y_{02},y_{03}) = (1.05,0.15,0)\;, \label{eq15}
\end{eqnarray}
respectively. The results are presented on {\bf Figure 1} - one can see that there exist
(in fact) the divergences between both trajectories. Next, we find the solutions for the
master system (\ref{eq8}) (the $x$-trajectory) and for its slave partner (\ref{eq9}) with active controllers
(\ref{eq12}) (the $y$-trajectory) for initial data (\ref{eq14}) and (\ref{eq15}) respectively. Now, we see that the
corresponding trajectories become synchronized - the vanishing in time error functions $e_i = y_i - x_i$ are presented on {\bf Figure 2}.
Additionally, we repeat the above numerical procedure for two another sets of initial data: $x_0 = (0,0.15,0.05)$ and $y_0=(0.05,0.05,0)$; the obtained results
are presented on {\bf Figures 3} and {\bf 4} respectively.

\section{Chaos synchronization of identical Sprott systems}

The used in pervious Section algorithm can be applied to the case of all remaining Sprott
systems as well. The obtained results are summarized in {\bf Table 2}, i.e., there are listed controllers
$u_1$, $u_2$ and $u_3$ for which the proper identical Sprott systems become synchronized
for arbitrary initial conditions $x_{01}$, $x_{02}$ and $x_{03}$ as well as $y_{01}$, $y_{02}$ and $y_{03}$.
However, as it was already mentioned in
Introduction, the control functions for Sprott-L and Sprott-M models have been provided
in paper \cite{[31]}. \\
\\
\\
\begin{tabular}{ccccc}
\hline $\;\;\;\;{\rm type}\;\;\;\;$ & $\;\;\;\;\;\;\;{\rm 1st\;controller}\;\;\;\;\;\;\;$ & $\;\;\;\;\;\;\;{\rm 2nd\;controller}\;\;\;\;\;\;\;$ & $\;\;\;\;\;\;\;{\rm 3rd\;controller}\;\;\;\;\;\;\;$\\
\hline ${\rm A}$ & $u_1=-(e_1+e_2)$ & $u_2 = e_1-e_2-y_2y_3+$ & $u_3=e_2(y_2+x_2)-e_3$\\
       $~~$ & $~~$ & $+x_2x_3$ & $~~$\\
       ${\rm B}$ & $u_1=x_2x_3-y_2y_3-e_1$ & $u_2=-e_1$ & $u_3=y_1y_2-x_1x_2-e_3$\\
       ${\rm C}$ & $u_1=x_2x_3-y_2y_3-e_1$ & $u_2=-e_1$ & $u_3=(x_1+y_1)e_1-e_3$\\
       ${\rm D}$ & $u_1=e_2-e_1$ & $u_2=-(e_1+e_2+e_3)$ & $u_3=x_1x_3-3(y_2+x_2)\cdot$\\
       $~~$ & $~~$ & $~~$ & $~\cdot e_2-e_3-y_1y_2$\\
       ${\rm E}$ & $u_1=x_1x_3-y_1y_3-e_1$ & $u_2=-(y_1+x_1)e_1$ & $u_3=4e_1-e_2$\\
       ${\rm F}$ & $u_1=-(e_1+e_2+e_3)$ & $u_2=e_1-1.5e_2$ & $u_3=-(y_1+x_1)e_1$\\
       ${\rm G}$ & $u_1=-(1.4e_1+e_3)$ & $u_2=x_1x_3-y_1y_3$ & $u_3=e_1-(e_2+e_3)$\\
       ${\rm H}$ & $u_1=e_2-(y_3+x_3)\cdot$ & $u_2=-(e_1+1.5e_2)$ & $u_3=-e_1$\\
       $~~$ & $\cdot e_3-e_1$ & $~~$ & $~~$\\
       ${\rm I}$ & $u_1=0.2e_2-e_1$ & $u_2=-(e_1+e_2+e_3)$ & $u_3=-(e_1+ (y_2+x_2)\cdot$\\
       $~~$ & $~~$ & $~~$ & $\cdot e_2)$\\
       ${\rm J}$ & $u_1=-(e_1+2e_3)$ & $u_2=e_2-e_3$ & $u_3=e_1-(1+y_2+x_3)\cdot$\\
       $~~$ & $~~$ & $~~$ & $\cdot e_2-e_3)$\\
       ${\rm K}$ & $u_1=e_3-e_1-y_1y_2+$ & $u_2=-e_1$ & $u_3=-(e_1+1.3e_3)$\\
       $~~$ & $+x_1x_2$ & $~~$ & $~~$\\
       ${\rm L}$ & $u_1=-(e_1+e_2+3.9e_3)$ & $u_2=-0.9(y_1+x_1)e_1$ & $u_3=e_1-e_3$\\
       \end{tabular}
       \begin{tabular}{ccccc}
$\;\;\;\;\;\;\;\;\;\;\;\;\;\;\;\;\;\;\;$ & $\;\;\;\;\;\;\;\;\;\;\;\;\;\;\;\;\;\;\;\;\;\;\;\;\;\;$ & $\;\;\;\;\;\;\;\;\;\;\;\;\;\;\;\;\;\;\;\;\;\;\;\;\;\;\;$ & $\;\;\;\;\;\;\;\;\;\;\;\;\;\;\;\;\;\;\;\;\;\;\;\;\;\;$\\
       ${\rm M}$ & $u_1=e_3-e_1$ & $u_2=-e_1(y_1+x_1)$ & $u_3=-(1.7e_1+e_2+e_3)$\\
       ${\rm N}$ & $u_1=2e_2-e_1$ & $u_2=-(e_1+(y_3+x_3)\cdot$ & $u_3=2e_1-e_2-e_3$\\
       $~~$ & $~~$ & $\cdot e_3+e_2)$ & $~~$\\
       ${\rm O}$ & $u_1=-(e_1+e_2)$ & $u_2=-(e_1+e_2)+e_3$ & $u_3=-(e_1+2.7e_2+$\\
       $~~$ & $~~$ & $~~$ & $+e_3)+x_1x_2-y_1y_2$\\
       ${\rm P}$ & $u_1=-(e_1+2.7e_2+$ & $u_2=e_1-(y_2+x_2)\cdot$ & $u_3=-(e_1+e_2+e_3)$\\
       $~~$ & $+e_3)$ & $\cdot e_2-e_2)$ & $~~$\\
       ${\rm Q}$ & $u_1=e_3-e_1$ & $u_2=e_1$ & $u_3=-(3.1e_1+(y_2+x_2)\cdot$\\
       $~~$ & $~~$ & $~~$ & $\cdot e_2)-1.5e_3$\\
       ${\rm R}$ & $u_1=-e_1+e_2$ & $u_2=-(e_2+e_3)$ & $u_3=x_1x_2-y_1y_2$\\
       ${\rm S}$ & $u_1=4e_2$ & $u_2=-(e_1+(y_3+x_3)\cdot$ & $u_3=-(e_1+e_3)$\\
       $~~$ & $~~$ & $\cdot e_3)-e_2$ & $~~$\\
\hline
\end{tabular}

\begin{center}
{\bf Table 2.} The active controllers for Sprott systems.
\end{center}

\section{Final remarks}
In this article we synchronize all identical Sprott systems defined in paper \cite{[10]} with
use of the active control method. Particularly, we find the corresponding so-called active
controllers listed in {\bf Table 2}. As an example we also study numerically synchronization
of Sprott-A model defined by the formulas (\ref{eq8}) and (\ref{eq9}).

It should be noted that the presented investigations can be extended in various ways.
For example, one may consider synchronization of Sprott models with use of others mentioned in
Introduction methods. The works in this direction already started and are in
progress.

\eject

\pagestyle{empty}
$~~~~~~~~~~~~~~~~~~$
\\
\\
\\
\\
\\
\\
\\
\\
\\
\begin{figure}[htp]
\includegraphics[width=\textwidth]{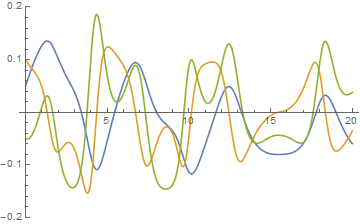}
\caption{The error functions $e_i = y_i - x_i$ for Sprott-A model defined by the system (\ref{eq8}) with the
initial conditions (\ref{eq14}) (the $x$-trajectory) and with the initial conditions (\ref{eq15}) (the $y$-trajectory). The blue line
corresponds to the $e_1$-error function, the orange one - to $e_2$ and the green one - to $e_3$ respectively.}\label{grysunek1}
\end{figure}
\begin{figure}[htp]
\includegraphics[width=\textwidth]{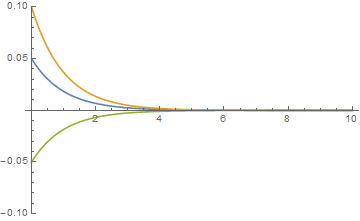}
\caption{The error functions $e_i = y_i - x_i$ for synchronized Sprott-A model defined by the master system (\ref{eq8}) with the
initial conditions (\ref{eq14}) (the $x$-trajectory) and by the slave system (\ref{eq9}) with the initial conditions (\ref{eq15}) (the $y$-trajectory).
The blue line corresponds to the $e_1$-error function, the orange one - to $e_2$ and the green one - to $e_3$ respectively.}\label{grysunek2}
\end{figure}
\begin{figure}[htp]
\includegraphics[width=\textwidth]{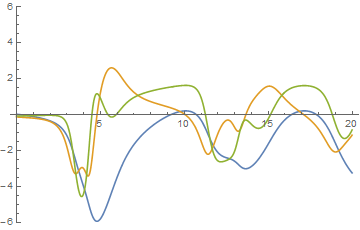}
\caption{The error functions $e_i = y_i - x_i$ for Sprott-A model defined by the system (\ref{eq8}) with the
initial conditions $x_{0} = (0,0.15,0.05)$ (the $x$-trajectory) and with the initial conditions $y_{0} = (0.05,0.05,0)$ (the $y$-trajectory). The blue line
corresponds to the $e_1$-error function, the orange one - to $e_2$ and the green one - to $e_3$ respectively.}\label{grysunek3}
\end{figure}
\begin{figure}[htp]
\includegraphics[width=\textwidth]{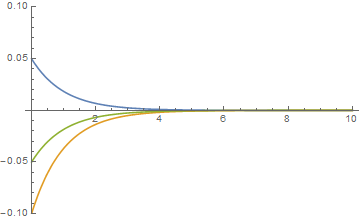}
\caption{The error functions $e_i = y_i - x_i$ for synchronized Sprott-A model defined by the master system (\ref{eq8}) with the
initial conditions $x_{0} = (0,0.15,0.05)$ (the $x$-trajectory) and by the slave system (\ref{eq9}) with the initial conditions $y_{0} = (0.05,0.05,0)$
(the $y$-trajectory). The blue line corresponds to the $e_1$-error function, the orange one - to $e_2$ and the green one - to $e_3$ respectively.}\label{grysunek4}
\end{figure}

\end{document}